\documentclass[10pt,journal]{IEEEtran}

\usepackage[final]{graphicx}
\usepackage{epsfig}
\usepackage{caption2}
\usepackage{amsmath}
\usepackage{amssymb,times}
\usepackage{bm}
\usepackage{cite}
\usepackage{color}
\usepackage[ruled,vlined]{algorithm2e}     

\newcommand{\ba}{\begin{array}}
\newcommand{\ea}{\end{array}}
\newcommand{\be}{\begin{displaymath}}
\newcommand{\ee}{\end{displaymath}}
\newcommand{\ben}{\begin{equation}}
\newcommand{\een}{\end{equation}}
\newcommand{\bena}{\begin{eqnarray}}
\newcommand{\eena}{\end{eqnarray}}
\newcommand{\beqa}{\begin{eqnarray*}}
\newcommand{\enqa}{\end{eqnarray*}}

\newcommand{\bc}{\begin{center}}
\newcommand{\ec}{\end{center}}
\newcommand{\bi}{\begin{itemize}}
\newcommand{\ei}{\end{itemize}}
\newcommand{\benu}{\begin{enumerate}}
\newcommand{\eenu}{\end{enumerate}}
\newcommand{\bdes}{\begin{description}}
\newcommand{\edes}{\end{description}}
\newcommand{\bt}{\begin{tabular}}
\newcommand{\et}{\end{tabular}}

\newcommand \thetabf{{\mbox{\boldmath$\theta$\unboldmath}}}

\newcommand \phibf{\mbox{\boldmath$\phi$\unboldmath}}

\newcommand \nbf{{\mathbf n}}

\newcommand \tbf{{\mathbf t}}

\newcommand \wbf{{\mathbf w}}
\newcommand \xbf{{\mathbf x}}
\newcommand \ybf{{\mathbf y}}
\newcommand \zbf{{\mathbf z}}

\newcommand \Bbf{{\mathbf B}}

\newcommand \Dbf{{\mathbf D}}
\newcommand \Ebf{{\mathbf E}}
\newcommand \Fbf{{\mathbf F}}
\newcommand \Gbf{{\mathbf G}}
\newcommand \Hbf{{\mathbf H}}
\newcommand \Ibf{{\mathbf I}}
\newcommand \Jbf{{\mathbf J}}

\newcommand \Pbf{{\mathbf P}}

\newcommand \Rbf{{\mathbf R}}
\newcommand \Sbf{{\mathbf S}}

\newcommand \Ubf{{\mathbf U}}
\newcommand \Vbf{{\mathbf V}}

\newcommand \Xbf{{\mathbf X}}

\newcommand{\circlambda}{\mbox{$\Lambda$
             \kern-.85em\raise1.5ex
             \hbox{$\scriptstyle{\circ}$}}\,}

\renewcommand \thetabf{\boldsymbol{\theta}}

\renewcommand \phibf{\boldsymbol{\phi}}

\IEEEoverridecommandlockouts

\begin{document}

\title{Detector Design and Performance Analysis for Target Detection in Subspace Interference}

\author{Weijian Liu, \emph{Senior Member, IEEE}, Jun~Liu, \emph{Senior Member, IEEE},  Tao Liu\\
Hui Chen, and Yong-Liang Wang, \emph{Member, IEEE}

\thanks{This work was partially supported by the National Natural Science Foundation of China (Nos. 62071482, 61871469, and 62171452).}
\thanks{W. Liu, H. Chen, and  Y.-L. Wang are with Wuhan Electronic Information Institute, Wuhan 430019, China (e-mail: liuvjian@163.com, chhglr@sina.com, and ylwangkjld@163.com).}
\thanks{J. Liu is with the Department of Electronic Engineering and Information Science, University of Science and Technology of China, Hefei 230027, China (e-mails: junliu@ustc.edu.cn).}
\thanks{T. Liu is with the School of Electronic Engineering, Naval University of Engineering, Wuhan 430033, China (e-mail: liutao1018@hotmail.com).}
}

\maketitle

\begin{abstract}
It is often difficult to obtain sufficient training data for adaptive signal detection, which is required to calculate the unknown noise covariance matrix. Additionally, interference is frequently present, which complicates the detecting issue. We provide a two-step method, termed interference cancellation before detection (ICBD), to address the issue of signal detection in the unknown Gaussian noise and subspace interference. The first involves projecting the test and training data to the interference-orthogonal subspace in order to suppress the interference. Utilizing traditional adaptive detector design ideas is the next stage. Due to the smaller dimension of the projected data, the ICBD-based detectors can function with little training data. The ICBD has two additional benefits over traditional detectors. Lower computational burden and proper operation with interference being in the training data are two additional benefits of ICBD-based detectors over conventional ones. We also give the statistical properties of the ICBD-based detectors and demonstrate their equivalence with the traditional ones in the special case of a large amount of training data containing no interference. \end{abstract}


\IEEEpeerreviewmaketitle
\section{Introduction}

Adaptive signal detection is a fundamental issue for signal processing, for which the noise covariance matrix is unknown and a set of training data is used to infer the noise covariance matrix \cite{TangRong19,RongAubry21,LiuLiu22SCIS}. Subspace interference is another common phenomenon that occurs in a known subspace but with unknown coordinates \cite{ScharfFriedlander94}. Numerous detectors, such those in \cite{BandieraBesson07,BandieraDeMaio07a,LiuLiu2015d, LiuLiu19TAESWald,SchniterByrne19,Fazlollahpoor20,YanAddabbo20,SunLiu22TAESInterf,SunLiu22TAESOrth}, were suggested for subspace interference and unknown noise. The articles above adopt two common presumptions. One is that there are sufficient training data. The other is that the interference only appears in the test data. However, in many situations, it could be challenging to obtain sufficient training data, and/or the interference may occupy the total range bins and hence exists in the test and training data \cite{KirsteinsTufts94,GauReed98}. In the first case, the sample covariance matrix (SCM) is singular, which
renders the detectors in \cite{BandieraBesson07,BandieraDeMaio07a,LiuLiu2015d, LiuLiu19TAESWald,SunLiu22TAESInterf,SunLiu22TAESOrth} invalid. The latter scenario lacks effective detectors for the three most used detector design criteria, namely, the generalized likelihood ratio test (GLRT), Wald test, and Rao test \cite{DeMaioKay10,LiuWang14,KayZhu16TSP}. This is the result of the too huge size of the unknown parameter space  \cite{AubryDeMaio14b}. 

In this letter, we address the challenge of signal detection when both the test and training data contain subspace interference and the number of training data too limited to form an invertible SCM. We introduce a simple and efficient solution to the problem. This work makes three main contributions. We first propose a two-step detection method. The test and training data are transformed using a semi-unitary matrix in the first stage to accomplish interference suppression. The columns of the semi-unitary matrix span an subspace that is orthogonal to the interference. The detectors are created in the second step utilizing detector design criteria of GLRT, Rao, and Wald test. Interference cancellation before detection (ICBD) is the name of the new method. Second, we demonstrate that when the interference only appears in the test data and there are enough training data, the GLRT, Rao test, and Wald test based on the ICBD are equivalent to the traditional ones. The ICBD methodology provides several advantages over traditional methods in this situation, including lesser processing cost brought on by the data's smaller dimension. Third, we derive the statistical distributions of the ICBD-based detectors and demonstrate that, in the situation of enough training data having no interference, the ICBD-based detectors are equivalent to the traditional detectors.

\section{Problem Formulation}
The following binary hypothesis test represents the detection problem to be solved:
\begin{equation}
\label{1}
\left\{ \begin{array}{l}
{\text{H}_0}:\xbf =  \Jbf\phibf  + \nbf,
{\kern 1pt} {\kern 1pt} {\kern 1pt} {\kern 1pt} {\kern 1pt} {\kern 1pt} {\kern 1pt}
{\xbf_l} = \Jbf\phibf_l+{\nbf_l},
{\kern 1pt} {\kern 1pt} {\kern 1pt} {\kern 1pt} {\kern 1pt} l = 1,\cdots,L,\\
{\text{H}_1}:\xbf = \Hbf\thetabf + \Jbf\phibf + \nbf,
{\kern 1pt} {\kern 1pt} {\kern 1pt} {\kern 1pt} {\kern 1pt} {\kern 1pt} {\kern 1pt}
{\xbf_l} = \Jbf\phibf_l+{\nbf_l},{\kern 1pt} {\kern 1pt} {\kern 1pt} {\kern 1pt} {\kern 1pt} l = 1,\cdots,L,
\end{array} \right.
\end{equation}
where under hypothesis $\text{H}_0$ the $N\times1$ test data vector $\xbf$ consists of noise $\nbf$ and subspace interference $\Jbf\phibf$, with the $N\times q$ full-column-rank matrix $\Jbf$ spanning the interference subspace, denoted as $<\Jbf>$, and the $q\times1$ vector $\phibf$ being the interference coordinate vector. Under hypothesis $\text{H}_1$, $\xbf$ contains signal $\Hbf\thetabf$, besides the noise and interference, with the $N\times p$ full-column-rank matrix $\Hbf$ spanning the signal subspace $<\Hbf>$, and the $p\times1$ vector $\thetabf$ being the signal coordinate vector.
$\Hbf$ and $\Jbf$ are known \emph{a priori}, whereas $\thetabf $ and $\phibf $ are unknown. Furthermore, the covariance matrix of $\nbf$, represented as $\Rbf_t$, is unknown under both hypotheses.
 A set of signal-free training data, represented as $\xbf_l$, $l=1,2,\cdots,L$, is required to estimate $\Rbf_t$, with $\nbf_l$ and $\phibf_l$ standing for the noise and interference's coordinate vectors of $\xbf_l$, respectively. In \eqref{1}, the interference in the test and training data share the same subspace. In practice, this can happen when the interference changes not very rapidly.

There are two reasons why there are no efficient conventional detectors for the detection problem in \eqref{1}. One is the presence of interference in the training data, and the other is the lack of sufficient training data (i.e., $L<N$) where the SCM is not invertible. Two cases of $\Rbf_t$ are considered in this letter. One is homogenous environment (HE), where ${\Rbf_t} = \Rbf$ \cite{Kelly86}, with $\Rbf$ denoting the noise covariance matrix of the training data. The other is partially homogeneous environment (PHE), where ${\Rbf_t} = {\sigma ^2}\Rbf$ \cite{ConteDeMaio01}, and $\sigma^2$ is an unidentified positive scaling factor, representing the power mismatch between the noise in training and test data.




\section{The Proposed ICBD}
\subsection{Derivation of the ICBD-based detectors}
The polar decomposition of the matrix $\Jbf$ is \cite[p.348]{Seber08}
\begin{equation}
\label{2}
\Jbf = {\Jbf_u}{\Dbf_\Jbf},
\end{equation}
where ${\Jbf_u} = \Jbf{({\Jbf^H}\Jbf)^{-\frac{1}{2}}}$ and ${\Dbf_\Jbf} = {({\Jbf^H}\Jbf)^{\frac{1}{2}}}$, with $(\cdot)^H$ being conjugate transpose. Note that $ < \Jbf >  = $ $ < {\Jbf_u} > $. There is an $N \times (N - q)$ matrix ${\Jbf_ \bot }$
 that makes the augmented matrix $\Ubf = [{\Jbf_u},{\Jbf_ \bot }]$  an $N \times N$ unitary matrix satisfying $\Jbf_ \bot ^H{\Jbf_ \bot } = {\Ibf_{N - q}}$ and $\Jbf_ \bot ^H\Jbf = {\mathbf{0}_{(N - q) \times q}}$. Pre-multiplying \eqref{1} by $\Jbf_ \bot ^H$, along with the fact $\Jbf_ \bot ^H\Jbf = {{\mathbf{0}}_{(N - q) \times q}}$, we have
\begin{equation}
\label{3}
\left\{ \begin{array}{l}
{\text{H}_0}:\ybf = \wbf,{\kern 1pt} {\kern 1pt} {\kern 1pt} {\kern 1pt} {\kern 1pt} {\kern 1pt} {\ybf_l} = {\wbf_l},{\kern 1pt} {\kern 1pt} {\kern 1pt} {\kern 1pt} {\kern 1pt} l = 1,2,...,L,\\
{\text{H}_1}:\ybf = {\Hbf_{{\Jbf_ \bot }}}\thetabf  + \wbf,{\kern 1pt} {\kern 1pt} {\kern 1pt} {\kern 1pt} {\kern 1pt} {\kern 1pt} {\ybf_l} = {\wbf_l},{\kern 1pt} {\kern 1pt} {\kern 1pt} {\kern 1pt} {\kern 1pt} l = 1,2,...,L,
\end{array} \right.
\end{equation}
where $\ybf = \Jbf_ \bot ^H\xbf$, ${\Hbf_{{\Jbf_ \bot }}} = \Jbf_ \bot ^H\Hbf$, $\wbf = \Jbf_ \bot ^H\nbf$, and ${\wbf_l} = \Jbf_ \bot ^H{\nbf_l}$. The covariance matrices of $\wbf$ and $\wbf_l$ are  ${\Rbf_{t{\Jbf_ \bot }}} \buildrel \Delta \over = \Jbf_ \bot ^H{\Rbf_t}{\Jbf_ \bot }  $ and ${\Rbf_{{\Jbf_ \bot }}}\buildrel \Delta \over =\Jbf_ \bot ^H\Rbf{\Jbf_ \bot }  $, respectively. 

Interestingly, the interference is rejected in both the test and training data \eqref{3}. Additionally, the detection issue in \eqref{3} has the same structure as those in \cite{RaghavanPulsone96,KrautScharf01,PastinaLombardo01,LiuXie14b}. Consequently, we can easily derive the GLRT, Rao test, and Wald test for the detection problem  in \eqref{3} using the results in \cite{RaghavanPulsone96,KrautScharf01,PastinaLombardo01,LiuXie14b}. The resulting GLRT, Rao test, and Wald test in HE are referred to as the ICBD-GLRT-HE, ICBD-Rao-HE, and ICBD-Wald-HE, respectively. Similarly, the GLRT, Rao, and Wald tests the PHE are denoted as the ICBD-GLRT-PHE, ICBD-Rao-PHE, and ICBD-Wald-PHE, respectively. Specifically, according to \cite{RaghavanPulsone96,PastinaLombardo01,LiuXie14b} the ICBD-GLRT-HE, ICBD-Rao-HE, and ICBD-Wald-HE are
\begin{equation}
\label{4}
t_{\text{ICBD-GLRT-HE}} = \frac {{{\tilde \ybf}^H}{\Pbf_{{{\tilde \Hbf}_{_{{\Jbf_ \bot }}}}}}\tilde \ybf}  {1 + {{\tilde \ybf}^H}\Pbf_{{{\tilde \Hbf}_{_{{\Jbf_ \bot }}}}}^ \bot \tilde \ybf},
\end{equation}
\begin{equation}
\label{Rao-HE}
t_\text{ICBD-Rao-HE}= 
\frac {{{\tilde \ybf}^H}{\Pbf_{{{\tilde \Hbf}_{_{{\Jbf_ \bot }}}}}}\tilde \ybf}  {(1+{\tilde \ybf}^H{\tilde \ybf})(1 + {{\tilde \ybf}^H}\Pbf_{{{\tilde \Hbf}_{_{{\Jbf_ \bot }}}}}^ \bot \tilde \ybf)},
\end{equation}
and
\begin{equation}
\label{5}
t_\text{ICBD-Wald-HE}^{} = {\tilde \ybf^H}{\Pbf_{{{\tilde \Hbf}_{_{{\Jbf_ \bot }}}}}}\tilde \ybf,
\end{equation}
respectively, where $\tilde \ybf = \Sbf_{{\Jbf_ \bot }}^{-\frac{1}{2}}\ybf$, ${\tilde \Hbf_{{\Jbf_ \bot }}} = \Sbf_{{\Jbf_ \bot }}^{-\frac{1}{2}}{\Hbf_{{\Jbf_ \bot }}}$, $\Pbf_{{{\tilde \Hbf}_{_{{\Jbf_ \bot }}}}}^ \bot  = {\Ibf_N} - {\Pbf_{{{\tilde \Hbf}_{_{{\Jbf_ \bot }}}}}}$, ${\Pbf_{{{\tilde \Hbf}_{_{{\Jbf_ \bot }}}}}} = {\tilde \Hbf_{{\Jbf_ \bot }}}{(\tilde \Hbf_{{\Jbf_ \bot }}^H{\tilde \Hbf_{{\Jbf_ \bot }}})^{ - 1}}\tilde \Hbf_{{\Jbf_ \bot }}^H$,
${\Sbf_{{\Jbf_ \bot }}} = \Jbf_ \bot ^H\Sbf{\Jbf_ \bot }$, $\Sbf =  {\Xbf_L}\Xbf_L^H$, and ${\Xbf_L} = [{\xbf_1},{\xbf_2},...,{\xbf_L}]$.
Analogously, using the results in \cite{KrautScharf01,DeMaioIommelli08,LiuXie14a}, the ICBD-GLRT-PHE, ICBD-Rao-PHE, and ICBD-Wald-PHE coincide with each other, given by
\begin{equation}
\label{6}
t_\text{ICBD-GLRT-PHE} = t_\text{ICBD-Rao-PHE}=t_\text{ICBD-Wald-PHE}= \frac{{{\tilde \ybf}^H}{\Pbf_{{{\tilde \Hbf}_{_{{\Jbf_ \bot }}}}}}\tilde \ybf} {{{\tilde \ybf}^H}\tilde \ybf}.
\end{equation}

It is important to note that $L$ is not too small. To ensure that $\Sbf_{\Jbf_{\bot}}$ is not singular, the condition $L\geq N-q$ must be satisfied.
\subsection{Statistical Properties of the ICBD-based Detectors}
Let
\begin{equation}
\label{8}
{\beta _\text{ICBD}} = {(1 + {\tilde \ybf^H}\Pbf_{{{\tilde \Hbf}_{_{{\Jbf_ \bot }}}}}^ \bot \tilde \ybf)^{ -1}},
\end{equation}
which is referred to as the loss factor for the ICBD-based detectors.
Using \eqref{4} and \eqref{8}, we can rewrite \eqref{Rao-HE}, \eqref{5} and \eqref{6} as
\begin{equation}
\label{Rao-HE2}
t_{\text{ICBD-Rao-HE}} = \frac{t_\text{ICBD-GLRT-HE}{\beta _\text{ICBD}}}{1+t_\text{ICBD-GLRT-HE}},
\end{equation}
\begin{equation}
\label{22}
t_\text{ICBD-Wald-HE} = \frac{t_{\text{ICBD-GLRT-HE}}} {{\beta _\text{ICBD}}},
\end{equation}
and
\begin{equation}
\label{GLRT-PHE}
t_{\text{ICBD-GLRT-PHE}} = \frac{t_\text{ICBD-GLRT-HE}} {1-\beta_\text{ICBD}+t_\text{ICBD-GLRT-HE}},
\end{equation}
respectively.



Using the results in \cite{RaghavanPulsone96,PastinaLombardo01}, we can show that the ICBD-GLRT-HE in \eqref{4}, conditioned on ${\beta _\text{ICBD}}$, is distributed as a complex noncentral F-distribution with $p$ and $L-N+q+1$ degrees of freedom (DOFs), and a noncentrality
\begin{equation}
\label{13}
{\rho _\text{ICBD}} = {\thetabf ^H}{\Hbf^H}{\Jbf_ \bot }{(\Jbf_ \bot ^H\Rbf{\Jbf_ \bot })^{ -1}}\Jbf_ \bot ^H\Hbf\thetabf,
\end{equation}
defined as the effective signal-to-noise ratio (eSNR).
The conditional distribution of ICBD-GLRT-HE under hypothesis $\text{H}_1$ is written symbolically as
\begin{equation}
\label{12}
t_{\text{ICBD-GLRT-HE}}\sim{\cal C}{{\cal F}_{p,L -N + q + 1}}({\rho _\text{ICBD}}{\beta _\text{ICBD}})  \ {\text{under}} \ {\text{H}_1},
\end{equation}
Under $\text{H}_0$, 
the statistical distribution of the ICBD-GLRT-HE reduces to 
\begin{equation}
\label{18}
t_{\text{ICBD-GLRT-HE}}\sim{\cal C}{{\cal F}_{p,L -N + q + 1}} \ \text{under}\ \text{H}_0 .
\end{equation}
Furthermore, we can quickly verify that the loss factor in \eqref{8} is distributed as a complex central Beta distribution with $L -N + p + q + 1$ and $N -p -q$ DOFs by applying the results in \cite{RaghavanPulsone96,PastinaLombardo01}, i.e.,
\begin{equation}
\label{19}
\beta _{\text{ICBD}} \sim{\cal C}{\cal B} _{L -N + p + q + 1,N -p -q} \ {\text{under}} \ {\text{H}_1} \ \text{and} \ {\text{H}_0}.
\end{equation}

Using the statistical distributions in \eqref{12}, \eqref{18}, and \eqref{19}, we can derive the expressions for the probability of detection (PD) and probability of false alarm (PFA) of the ICBD-GLRT-HE. Additionally, we can obtain the PDs and PFAs of the ICBD-Rao-HE, ICBD-Wald-HE, and ICBD-GLRT-PHE according to the statistical dependence in \eqref{Rao-HE2}-\eqref{GLRT-PHE}, as well as the statistical distributions in \eqref{12}, \eqref{18}, and \eqref{19}.
Due to space restrictions, however, detailed expressions for the PDs and PFAs are not provided here.

Except the noncertainty parameter eSNR, it appears that the conditional distribution of the ICBD-GLRT-HE in \eqref{4} is the same as the GLRT-HE for \eqref{1} when the training data number is sufficient and the interference only occurs in the test data. In fact, the eSNR for the ICBD-GLRT-HE in \eqref{13} is equivalent to the eSNR for the GLRT-HE, given as
\begin{equation}
\label{14}
{\rho _\text{CON}} = {\thetabf ^H}{\bar \Hbf^H}\Pbf_{\bar \Jbf}^ \bot \bar \Hbf\thetabf,
\end{equation}
where $\bar\Hbf=\Rbf^{-\frac{1}{2}}\Hbf$, $\Pbf_{\bar \Jbf}^ \bot=\Ibf_N- \Pbf_{\bar \Jbf}$, $\Pbf_{\bar \Jbf}=\bar \Jbf(\bar \Jbf^H\bar \Jbf)^{-1}\bar \Jbf^H$, and $\bar\Jbf=\Rbf^{-\frac{1}{2}}\Jbf$.
The proof of the equality of \eqref{13} and \eqref{14} is given in Appendix A.

In the situation that adequate training data exist and interference only lies in the test data, the ICBD-based detectors developed using the GLRT, Rao test, and Wald test are actually equivalent to the corresponding conventional detectors. The next subsection addresses this fact.
%

\subsection{Equivalence of the ICBD-based Detectors and Conventional Detectors When  Training Data Are Sufficient and Interference Only Exists in the Test Data}

When the detection problem in \eqref{1} is modified to the case that $\xbf_l=\nbf_l$ and $L \ge N$, then the GLRT, Rao test, and Wald test in HE are \cite{BandieraDeMaio07a,SunLiu22TAESInterf}
\begin{equation}
\label{44}
t_\text{GLRT-HE} = \frac{{{\tilde \xbf}^H}{\Pbf_{\Pbf_{\tilde \Jbf}^ \bot \tilde \Hbf}}\tilde \xbf}  {1 + {{\tilde \xbf}^H}\Pbf_{\tilde \Jbf}^ \bot \tilde \xbf -{{\tilde \xbf}^H}{\Pbf_{\Pbf_{\tilde \Jbf}^ \bot \tilde \Hbf}}\tilde \xbf},
\end{equation}
\begin{equation}
\label{RaoHE}
t_\text{Rao-HE} = \frac{{{\tilde \xbf}^H}{\Pbf_{\Pbf_{\tilde \Jbf}^ \bot \tilde \Hbf}}\tilde \xbf}  {({1 + {{\tilde \xbf}^H}\Pbf_{\tilde \Jbf}^ \bot \tilde \xbf}) ({1 + {{\tilde \xbf}^H}\Pbf_{\tilde \Bbf}^ \bot \tilde \xbf}) },
\end{equation}
and
\begin{equation}
\label{40}
{t_\text{Wald-HE}} = {\tilde \xbf^H}{\Pbf_{\Pbf_{\tilde \Jbf}^ \bot \tilde \Hbf}}\tilde \xbf,
\end{equation}
respectively, where $\tilde \xbf = {\Sbf^{ -\frac{1}{2}}}\xbf$,  $\tilde \Jbf = {\Sbf^{-\frac{1}{2}}}\Jbf$,
$\Pbf_{\tilde \Jbf}^ \bot  = {\Ibf_N} -{\Pbf_{\tilde \Jbf}}$,
${\Pbf_{\tilde \Jbf}} = \tilde \Jbf{({\tilde \Jbf^H}\tilde \Jbf)^{ - 1}}{\tilde \Jbf^H} $,
$\tilde \Bbf = {\Sbf^{ -\frac{1}{2}}}\Bbf$, $\Bbf = [\Hbf,\Jbf]$,
 ${\Pbf_{\tilde \Bbf}} = \tilde \Bbf{({\tilde \Bbf^H}\tilde \Bbf)^{ - 1}}{\tilde \Bbf^H}$,
 $\Pbf_{\tilde \Bbf}^ \bot  = {\Ibf_N} -
 {\Pbf_{\tilde \Bbf}}$,
 $\tilde \Hbf = {\Sbf^{ -\frac{1}{2}}}\Hbf$, and ${\Pbf_{\Pbf_{\tilde \Jbf}^ \bot \tilde \Hbf}} = \Pbf_{\tilde \Jbf}^ \bot \tilde \Hbf{({\tilde \Hbf^H}\Pbf_{\tilde \Jbf}^ \bot \tilde \Hbf)^{ -1}}{\tilde \Hbf^H}\Pbf_{\tilde \Jbf}^ \bot $.
The GLRT, Rao test, and Wald test in PHE for the problem in \eqref{1} with $\xbf_l=\nbf_l$ and $L \ge N$ coincide with each other, found to be \cite{SunLiu22TAESInterf}
\begin{equation}
\label{45}
t_\text{GLRT-PHE} = t_\text{Rao-PHE} = t_\text{Wald-PHE} = \frac{{{\tilde \xbf}^H}{\Pbf_{\Pbf_{\tilde \Jbf}^ \bot \tilde \Hbf}}\tilde \xbf}  {{{\tilde \xbf}^H}\Pbf_{\tilde \Jbf}^ \bot \tilde \xbf},
 \end{equation}

To demonstrate that ICBD-based detectors and  conventional detectors are equivalent when $\xbf_l=\nbf_l$ and $L \ge N$, we need to show the equalities of \eqref{4}-\eqref{6} and \eqref{44}-\eqref{45}, respectively. Detailed proof is shown in Appendix B.


When the interference matrix $\Jbf$ is obtained by electronic support measures (ESM) in advance, the main computational complexity for the ICBD-based detectors is the matrix inversion with a dimension of $(N-q)\times(N-q)$, while the main computational complexity for the conventional detectors  in \eqref{44}-\eqref{45} is the matrix inversion with a dimension of $N\times N$. Hence, the ICBD-based detectors have lower computational complexity than the conventional detectors. 
Furthermore, the ICBD-based detectors are able to operate when the number of the training data is less than the dimension of the test data, i.e., $L < N$, whereas the conventional detectors in \eqref{44}-\eqref{45} cannot due to the singularity of the conventional SCM $\Sbf$. However, the ICBD-based detectors require $L \ge N - q$ to ensure that the transformed SCM ${\Sbf_{{\Jbf_ \bot }}}$ is nonsingular. 

\section{Numerical examples}
We provide a numerical example to illustrate the detection performance of the proposed detectors in the sample-starved scenario. For simplicity, we only consider the case of HE.
To demonstrate the detection performance of the proposed detectors in the sample-starved scenario, we give a numerical example. For Monte Carlo simulations, we set $\Rbf(i,j)=\epsilon^{|i-j|^2}$, and  $100/\text{PFA}$ and $10^4$ independent trials are carried out to calculate the PFA and PD, respectively. The remaining parameters are: $N=24$, $L=23$, $p=2$, $q=6$, $\sin^2\psi=0.9$, $\text{cos}^2\vartheta=1$, $\epsilon=0.95$, and $\text{PFA}=10^{-4}$, where $\sin^2\psi$ and $\text{cos}^2\vartheta $ are defined in (73) and (74), respectively, in \cite{LiuLiu18AES}.
We choose  $\thetabf=\thetabf_0 a$, $\thetabf_0$ and $\phibf$ are randomly generated. However, after generation, $\thetabf_0$ and $\phibf$ are fixed. The constant $a$  is chosen according to a given eSNR.

For theoretical results and simulation results, respectively, the legends ``TH'' and ``MC'' are used. The results in Fig. 1 demonstrates that there is good agreement between theoretical findings and Monte Carlo simulations.
The ICBD-GLRT-HE offers the best detection performance of all the detector.
In Fig. 1, two intriguing discoveries are observed. One is that the ICBD-GLRT-PHE can offer higher PD than the ICBD-Wald-HE for the eSNR less than 23.7 dB. The other is that the PD of the ICBD-Rao-HE is extremely low.

\begin{figure}
	\centering
		\includegraphics[width=0.5\textwidth]{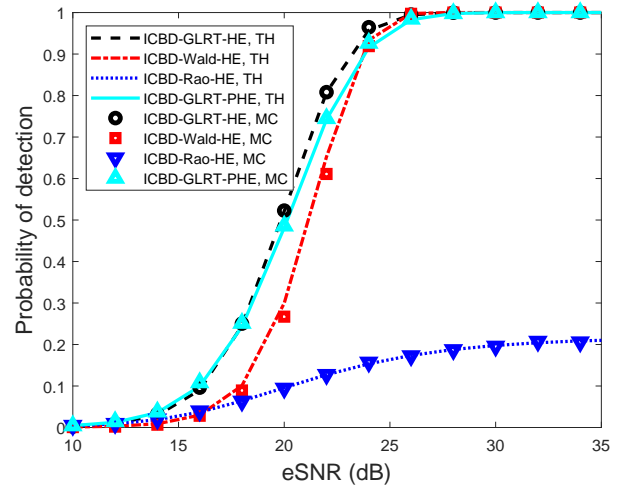}\\
	\caption{PDs of the proposed detectors under different eSNRs.}
	\label{Fig_ICBDTHMC} 
\end{figure}

\section{Conclusions}
With limited training data that contain interference, we presented the ICBD technique for detecting a signal in interference and noise. The interference suppression and conventional adaptive detector design make up the two fundamental components of the ICBD technique. In the special scenario of sufficient training data having no interference, the ICBD-GLRT-HE, ICBD-Rao-HE, ICBD-Wald-HE, and ICBD-GLRT-PHE are equivalent to the GLRT-HE, Rao-HE, Wald-HE, and GLRT-PHE, respectively. However, in the general scenario of limited training data, the ICBD-based detectors can work properly.
Additionally, the ICBD-based detectors are less computationally demanding than the traditional detectors in the above special scenario. Numerical examples show that the ICBD-GLRT-HE achieves the highest PD. However, to ensure that the transformed SCM is not singular, the number of training data should no less than $N-q$.

Possible future work may be generalization of the results to the case of distributed target or the case of inaccurate knowledge about the interference subspace.

\appendices 

\section{Proof of the Equality of \eqref{13} and \eqref{14}}
To prove the equality of \eqref{13} and \eqref{14}, we need to show
\begin{equation}
\label{71}
{\Rbf^{ -
1}} -{\Rbf^{ -1}}{\Jbf_u}{(\Jbf_u^H{\Rbf^{ -1}}{\Jbf_u})^{ - 1}}\Jbf_u^H{\Rbf^{ -
1}} = {\Jbf_ \bot }{(\Jbf_ \bot ^H\Rbf{\Jbf_ \bot })^{ -1}}\Jbf_ \bot ^H,
\end{equation}
where ${\Jbf_u} = \Jbf{({\Jbf^H}\Jbf)^{ -\frac{1}{2}}}$. One can easily verify 
\begin{equation}
\label{72}
{\Ubf^H}{\Jbf_u} = {\Ebf_q} = {[{\Ibf_q},{\mathbf{0}_{q \times (N -q)}}]^T},
\end{equation}
\begin{equation}
\label{73}
{\Ubf^H}{\Jbf_v} = \Vbf_{N -q}^{} = {[{\mathbf{0}_{(N -q) \times q}},{\Ibf_{N -q}}]^T},
\end{equation}
where $\Ubf = [{\Jbf_u},{\Jbf_ \bot }]$. 
Pre- and post-multiplying \eqref{71} by $\Ubf^H$ and $\Ubf$, respectively, and using \eqref{72} and \eqref{73}, results in
\begin{equation}
\label{74}
\begin{array}{l}
\Rbf_\Ubf^{ -1} -
\Rbf_\Ubf^{ -1}{\Ebf_q}{(\Ebf_q^H\Rbf_\Ubf^{ -1}{\Ebf_q})^{ - 1}}\Ebf_q^H\Rbf_\Ubf^{ -1} \\
{\kern 1pt} {\kern 1pt} {\kern 1pt} {\kern 1pt} {\kern 1pt} {\kern 1pt} {\kern 1pt} {\kern 1pt} {\kern 1pt} {\kern 1pt} {\kern 1pt} {\kern 1pt} {\kern 1pt} {\kern 1pt} {\kern 1pt} {\kern 1pt} {\kern 1pt} {\kern 1pt} {\kern 1pt} {\kern 1pt} {\kern 1pt} {\kern 1pt}
= \Vbf_{N -q}^{}{(\Vbf_{N - q}^H{\Rbf_\Ubf}\Vbf_{N -q}^{})^{ -1}}\Vbf_{N -q}^H,
\end{array}
\end{equation}
where $\Rbf_\Ubf^{} = {\Ubf^H}\Rbf\Ubf$. According to \eqref{73}, the right-hand side of \eqref{74} can be further expressed as
\begin{equation}
\label{75}
\Vbf_{N -q}^{}{(\Vbf_{N -q}^H{\Rbf_\Ubf}\Vbf_{N -q}^{})^{ -1}}\Vbf_{N -q}^H =\Gbf,
\end{equation}
where $\Gbf=\left[{{\mathbf{0}_{q,q}}}, \ {{\mathbf{0}_{q,N -
q}}};
{{\mathbf{0}_{N -q,q}}}, \ {\Rbf_{22}^{ -1}}\right]$,
$\Rbf_{22}$ are the last $N-q$ rows and $N-q$ columns of $\Rbf_\Ubf$. According to the partitioned matrix inversion formula \cite[p.1350]{VanTrees02}, we have
\begin{equation}
\label{76}
\Rbf_\Ubf^{ -1} -\Rbf_\Ubf^{ -1}{\Ebf_q}{(\Ebf_q^H\Rbf_\Ubf^{ -
1}{\Ebf_q})^{ -1}}\Ebf_q^H\Rbf_\Ubf^{ -1} =\Gbf.
\end{equation}
Consequently, \eqref{71} follows. The proof is now complete.
\section{Proof of the Equalities of \eqref{4}-\eqref{6} and \eqref{44}-\eqref{45}}
We successively prove the equalities of \eqref{5} and \eqref{40}, \eqref{4} and \eqref{44}, \eqref{Rao-HE} and \eqref{RaoHE}, \eqref{6} and \eqref{45}.
To prove the quality of \eqref{5} and \eqref{40}, we need to show
\begin{equation}
\label{46}
	{\Sbf^{ -\frac{1}{2}}}{\Pbf_{\Pbf_{\tilde \Jbf}^ \bot \tilde \Hbf}}{\Sbf^{-\frac{1}{2}}} = {\Jbf_ \bot }\Sbf_{{\Jbf_ \bot }}^{-\frac{1}{2}}{\Pbf_{{{\tilde \Hbf}_{_{{\Jbf_ \bot }}}}}}\Sbf_{{\Jbf_ \bot }}^{-\frac{1}{2}}\Jbf_ \bot ^H.
 \end{equation}
To this end, we first prove the following equation
\begin{equation}
\label{47}
 \text{Ker}(\mathord{\buildrel{\lower3pt\hbox{$\scriptscriptstyle\frown$}}
\over \Jbf} _ \bot ^H) =  < \tilde \Jbf >,
\end{equation}
where ${\mathord{\buildrel{\lower3pt\hbox{$\scriptscriptstyle\frown$}}
\over \Jbf} _ \bot } = {\Sbf^{\frac{1}{2}}}{\Jbf_ \bot }$, $\text{Ker}<\cdot>$ denotes the null space. Notice that if $< {\mathord{\buildrel{\lower3pt\hbox{$\scriptscriptstyle\frown$}}
\over \Jbf} _ \bot }{ > ^ \bot } =  < \tilde \Jbf >$, that is, $< {\Sbf^{\frac{1}{2}}}{\Jbf_ \bot }{ > ^ \bot } =  < {\Sbf^{ -\frac{1}{2}}}\Jbf >$, then \eqref{47} follows, with $<\cdot>^\bot$ denoting the orthogonal complement.
Let $\tbf \in  < {\Sbf^{ -\frac{1}{2}}}\Jbf > $ and $\zbf \in  < {\Sbf^{\frac{1}{2}}}{\Jbf_ \bot } > $. Then there exist two vectors $\xbf_1$ and $\xbf_2$ such that $\tbf = {\Sbf^{ -\frac{1}{2}}}\Jbf{\xbf_1}$ and $\zbf = {\Sbf^{\frac{1}{2}}}{\Jbf_ \bot }{\xbf_2}$. As a result, ${\tbf^H}\zbf = \xbf_1^H{\Jbf^H}{\Sbf^{-\frac{1}{2}}}{\Sbf^{\frac{1}{2}}}{\Jbf_ \bot }{\xbf_2} = 0$. Thus, $< {\Sbf^{ -\frac{1}{2}}}\Jbf >  \subseteq  < {\Sbf^{\frac{1}{2}}}{\Jbf_ \bot }{ > ^ \bot }$.
Conversely, let $\zbf \in  < {\Sbf^{\frac{1}{2}}}{\Jbf_ \bot }{ > ^ \bot }$, then ${\zbf^H}{\Sbf^{\frac{1}{2}}}{\Jbf_ \bot } =\mathbf{0}_{(N -q) \times 1}^T$. Define $\Fbf = {\Sbf^{ -\frac{1}{2}}}[\Jbf,{\Jbf_ \bot }]$, which 
can span the entire space. It follows that there exist two column vectors $\xbf_3$ and $\xbf_4$ such that  $\zbf = \Fbf{[\xbf_3^T,\xbf_4^T]^T} = {\Sbf^{-\frac{1}{2}}}\Jbf{\xbf_3} + {\Sbf^{ -\frac{1}{2}}}{\Jbf_ \bot }{\xbf_4}$. Pre-multiplying the identity above by $\Jbf_ \bot ^H{\Sbf^{\frac{1}{2}}}$, along with ${\zbf^H}{\Sbf^{\frac{1}{2}}}{\Jbf_ \bot } = \mathbf{0}_{(N -q) \times 1}^T$ and ${\Jbf^H}{\Jbf_ \bot } = {\mathbf{0}_{q \times (N -q)}}$, results in $\Jbf_ \bot ^H{\Jbf_ \bot }{\xbf_4} = {\mathbf{0}_{(N -q) \times 1}}$. Hence, ${\xbf_4} = {\mathbf{0}_{(N -
q) \times 1}}$ and  $\zbf = {\Sbf^{ -\frac{1}{2}}}\Jbf{\xbf_3}$. Therefore, $< {\Sbf^{\frac{1}{2}}}{\Jbf_ \bot }{ > ^ \bot } \subseteq  < {\Sbf^{ -\frac{1}{2}}}\Jbf > $. This completes the proof of \eqref{47}.
It follows from \eqref{47} that
\begin{equation}
\label{48}
\Pbf_{\mathord{\buildrel{\lower3pt\hbox{$\scriptscriptstyle\frown$}}
\over \Jbf} _ \bot} = \Pbf_{\tilde \Jbf}^ \bot.
\end{equation}
Hence, the left-hand side of \eqref{46} can be expanded as
\begin{equation}
\label{49}
\begin{array}{c}
\begin{aligned}
{\Sbf^{ -\frac{1}{2}}}&{\Pbf_{\Pbf_{\tilde \Jbf}^ \bot \tilde \Hbf}}{\Sbf^{-\frac{1}{2}}} = {\Sbf^{ -\frac{1}{2}}}{\Pbf_{{{{\bf{\mathord{\buildrel{\lower3pt\hbox{$\scriptscriptstyle\frown$}}
\over J} }}}_ \bot }}}{\Sbf^{ -\frac{1}{2}}}\Hbf\\
\qquad \quad\cdot &
{({\Hbf^H}{\Sbf^{ -\frac{1}{2}}}{\Pbf_{{{{\bf{\mathord{\buildrel{\lower3pt\hbox{$\scriptscriptstyle\frown$}}
\over J} }}}_ \bot }}}{\Sbf^{-\frac{1}{2}}}\Hbf)^{ -
1}} {\Hbf^H}{\Sbf^{ -\frac{1}{2}}}{\Pbf_{{{{\bf{\mathord{\buildrel{\lower3pt\hbox{$\scriptscriptstyle\frown$}}
\over J} }}}_ \bot }}}{\Sbf^{ -\frac{1}{2}}}.
\end{aligned}
\end{array}
\end{equation}
The right-hand side of \eqref{46} can be recast as
\begin{equation}
\label{50}
\begin{array}{l}
{\Jbf_ \bot }\Sbf_{{\Jbf_ \bot }}^{-\frac{1}{2}}{\Pbf_{{{\tilde \Hbf}_{_{{\Jbf_ \bot }}}}}}\Sbf_{{\Jbf_ \bot }}^{-\frac{1}{2}}\Jbf_ \bot ^H = {\Jbf_ \bot }{(\mathord{\buildrel{\lower3pt\hbox{$\scriptscriptstyle\frown$}}
\over \Jbf} _ \bot ^H{{\mathord{\buildrel{\lower3pt\hbox{$\scriptscriptstyle\frown$}}
\over \Jbf} }_ \bot })^{ -1}}\Jbf_ \bot ^H\Hbf\\
{\kern 1pt} {\kern 1pt} {\kern 1pt} {\kern 1pt} {\kern 1pt} {\kern 1pt} {\kern 1pt} {\kern 1pt} {\kern 1pt} {\kern 1pt} {\kern 1pt} {\kern 1pt} {\kern 1pt}  \cdot
{[{\Hbf^H}{\Jbf_ \bot }{(\mathord{\buildrel{\lower3pt\hbox{$\scriptscriptstyle\frown$}}
\over \Jbf} _ \bot ^H{{\mathord{\buildrel{\lower3pt\hbox{$\scriptscriptstyle\frown$}}
\over \Jbf} }_ \bot })^{ -1}}\Jbf_ \bot ^H\Hbf]^{ -
1}}{\Hbf^H}{\Jbf_ \bot }{(\mathord{\buildrel{\lower3pt\hbox{$\scriptscriptstyle\frown$}}
\over \Jbf} _ \bot ^H{{\mathord{\buildrel{\lower3pt\hbox{$\scriptscriptstyle\frown$}}
\over \Jbf} }_ \bot })^{ -1}}\Jbf_ \bot ^H.
\end{array}
\end{equation}
It follows that \eqref{50} is equal to \eqref{49} due to the definition of ${\mathord{\buildrel{\lower3pt\hbox{$\scriptscriptstyle\frown$}}
\over \Jbf} _ \bot }$, given in the line below \eqref{47}. This completes the proof of 
the equality of \eqref{40} and \eqref{5}.

Next, we show the equality of \eqref{4} and \eqref{44}. Pre- and post-multiplying \eqref{46} with $\xbf^H$ and $\xbf$, respectively, yields
\begin{equation}
\label{51}
{\tilde \ybf^H}{\Pbf_{{{\tilde \Hbf}_{_{{\Jbf_ \bot }}}}}}\tilde \ybf = {\tilde \xbf^H}{\Pbf_{\Pbf_{\tilde \Jbf}^ \bot \tilde \Hbf}}\tilde \xbf.
\end{equation}
Moreover, according to the definition of $\ybf$, we have
\begin{equation}
\label{52}
\begin{array}{c}
\begin{aligned}
{{\tilde \ybf}^H}\tilde \ybf &= {\xbf^H}{\Jbf_ \bot }{(\Jbf_ \bot ^H\Sbf{\Jbf_ \bot })^{ -1}}\Jbf_ \bot ^H\xbf\\
 &= {{\tilde \xbf}^H}{\Sbf^{\frac{1}{2}}}{\Jbf_ \bot }{(\Jbf_ \bot ^H\Sbf{\Jbf_ \bot })^{ -1}}\Jbf_ \bot ^H{\Sbf^{\frac{1}{2}}}\tilde \xbf\\
 &= {{\tilde \xbf}^H}\Pbf_{\tilde \Jbf}^ \bot \tilde \xbf,
\end{aligned}
\end{array}
\end{equation}
where we have used \eqref{48}. Subtracting \eqref{52} from \eqref{51} leads to
\begin{equation}
\label{53}
{\tilde \ybf^H}\Pbf_{{{\tilde \Hbf}_{_{{\Jbf_ \bot }}}}}^ \bot \tilde \ybf = {\tilde \xbf^H}\Pbf_{\tilde \Jbf}^ \bot \tilde \xbf -
{\tilde \xbf^H}{\Pbf_{\Pbf_{\tilde \Jbf}^ \bot \tilde \Hbf}}\tilde \xbf.
\end{equation}
This completes the equality of \eqref{4} and \eqref{44}.

The equality of \eqref{Rao-HE} and \eqref{RaoHE} can be proved using \eqref{46} and \eqref{52}, while the equality of \eqref{6} and \eqref{45} can be proved using \eqref{46}, \eqref{52}, and \eqref{53}.

{\small
\bibliographystyle{IEEEtran}
\bibliography{D:/LaTexReference/Detection}
}

\end{document}